\documentclass[aps, prb, twocolumn, dvipsnames,  superscriptaddress,shortbibliography]{revtex4-1}
\usepackage{graphicx, xcolor}
\usepackage{tabularx}
\usepackage{tablefootnote}
\usepackage{float}
\usepackage{amsmath} 
\usepackage{hyperref}
\usepackage{mathtools}
\usepackage{array, cellspace, hhline}
\hypersetup{
        colorlinks,
   citecolor=blue,
   filecolor=blue,
   linkcolor=blue,
   urlcolor=blue,
   breaklinks=true}

\usepackage[utf8]{inputenc}
\usepackage{braket}

\newcommand\ColorSection[1]{{\textcolor{RoyalBlue}{\textit{#1 --- }}}}  

\begin{document}
\title{Flat bands in bilayer graphene induced by proximity with polar $h$-BN superlattices}
\author{Marta Brzezi\'{n}ska}
\author{Oleg V. Yazyev}
\affiliation{Institute of Physics, Ecole Polytechnique Fédérale de Lausanne (EPFL), CH-1015 Lausanne, Switzerland}

\date{\today}

\begin{abstract}
Motivated by the observation of polarization superlattices in twisted multilayers of hexagonal boron nitride ($h$-BN), we address the possibility of using these heterostructures for tailoring the properties of multilayer graphene by means of the electrostatic proximity effect. 
By using the combination of first-principles and large-scale tight-binding model calculations coupled via the Wannier function approach, we demonstrate the possibility of creating a sequence of well-separated flat-band manifolds in AB-stacked bilayer graphene at experimentally relevant superlattice periodicities above $\sim$30~nm. Our calculations show that the details of band structures depend on the local inversion symmetry breaking and the vertical electrical polarization, which are directly related to the atomic arrangement.
The results advance the atomistic characterization of graphene-based systems in a superlattice potential beyond the continuum model.
\end{abstract}
\maketitle
\ColorSection{Introduction}
Van der Waals (vdW) heterostructures based on two-dimensional (2D) materials manifest dramatically different electronic, optical, and structural properties as opposed to their parent layered materials~\cite{Novoselov2016, Liu2016, Duong2017}.
More recently, moir\'e superlattices, with twisted bilayer graphene (TBG) being the first and the most prominent example, were shown to exhibit a wide range of exotic phenomena including unconventional superconductivity~\cite{Cao2018, Isobe2018, Wu2018, Po2018, Yankowitz2019, Lian2019, Saito2020, Oh2021, Lothman2022} and the Chern insulator phase~\cite{Cao2018a, Nuckolls2020, Serlin2020, Ledwith2020, Repellin2020, Stepanov2021, Wu2021, Pierce2021, Saito2021}.
Hexagonal boron nitride ($h$-BN) is usually employed as an encapsulating component in devices based on graphene heterostructures thanks to the closely matching crystal structure and large band gap of this material~\cite{Dean2010, Xue2011, Dean2012, Wang2017}.
Bulk $h$-BN is a layered material characterized by the AA' stacking configuration of individual honeycomb lattice layers: boron (nitrogen) atoms in one layer oppose nitrogen (boron) atoms in the two adjacent layers, resulting in a crystal with inversion symmetry~\cite{Constantinescu2013}.
However, polar $h$-BN heterostructures realizing stacking sequences with no inversion symmetry can be engineered.
Very recently, it was shown that interfacial ferroelectricity in such artificial Bernal-stacked (AB configuration) $h$-BN bilayers can be controlled by external electric field and the lateral shift of one of the layers~\cite{Woods2021, Yasuda2021, Stern2021}.
This unconventional sliding mechanism for the manipulation of the out-of-plane electrical polarization has been identified in heterostructures based on other vdW materials such as transition metal dichalcogenides~\cite{Fei2018, Wan2022, Wang2022} and persists at room temperature~\cite{He2022}, holding promise for atomically thin transistors~\cite{Wu2021a, Deb2022, Koprivica2022, Wang2023}.
These polar heterostructures typically exhibit a superlattice of triangular domains with alternating vertical polarization of magnitude $P_z$ due to a small residual twist accompanied by lattice relaxation.
Such a staggered polarization results a superlattice potential that can significantly affect the electronic structure of a system.
For instance, it has been demonstrated that $h$-BN bilayers can be used to induce ferroelectricity in BLG, which has no electric order in its initial state~\cite{Zheng2020}.
While several general frameworks have recently been proposed to recognize the origin and quantify 2D ferroelectricity (using group theory~\cite{Ji2022} or local registry index~\cite{Cao2022}), a comprehensive understanding of the connection between atomic environment, polarization, and electronic properties in these complex heterostructures is still missing.

\begin{figure}[!htbp]
    \centering
    \includegraphics[width=0.95\columnwidth]{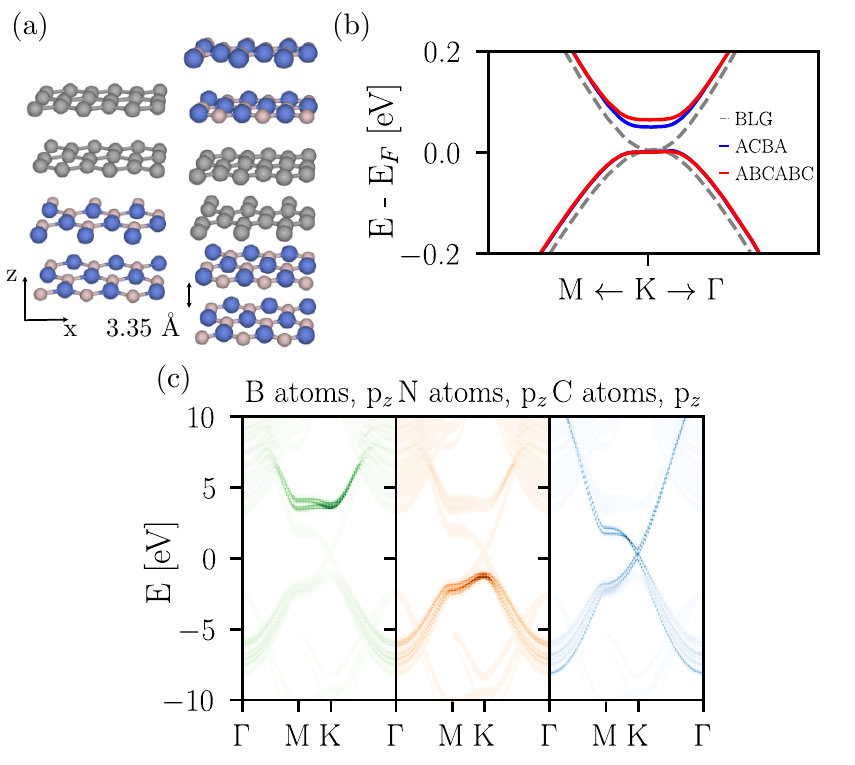}
    \caption{Non-centrosymmetric graphene/$h$-BN heterostructures and their electronic properties. (a) Schematic representation of two representative stacking configurations, ACBA and ABCABC, where BLG is deposited on or encapsulated between polar $h$-BN  bilayers. (b) Dispersion relation around the K point for pristine BLG, ACBA 2$h$-BN/BLG, and ABCABC 2$h$-BN/BLG/2$h$-BN. In the presence of $h$-BN, BLG develops a gap, the value of which depends on the details of the structure. (c) Projected band structures for the ABCABC configuration. The bands at the Fermi level are composed exclusively of the $p_z$ orbitals of C atoms.}
    \label{fig:dft_results}
\end{figure}

In this work, we show that polarization superlattices based on $h$-BN can be used to induce flat bands in AB-stacked bilayer graphene (BLG) with no twist whatsoever via the electrostatic proximity effect. More specifically, we perform a systematic study of all possible non-centrosymmetric BLG/$h$-BN heterostructures, in which the graphene component is either deposited on polar $h$-BN bilayers (denoted 2$h$-BN/BLG) or sandwiched between them (2$h$-BN/BLG/2$h$-BN) [cf. Fig.~\ref{fig:dft_results}(a)].
By combining first-principles calculations with the Wannier functions formalism, we address the gap opening mechanisms in the BLG spectrum.
To substantiate our results, we construct a tight-binding (TB) model incorporating alternating electrostatic potential distribution constructed from the on-site terms obtained via the wannierization procedure.
The TB calculations performed on large heterostructures show that for experimentally relevant superlattice periodicities $L>30$~nm, multiple of flat-band manifolds separate from the remote bands. 
Our results go beyond the band structure engineering using a superlattice potential in continuum models~\cite{Park2008, Barbier2010, Dubey2013, Forsythe2018, Shi2019, Huber2020, Li2021, Yang2022, GarciaRuiz2021, GarciaRuiz2023, Park2023}, including BLG~\cite{Killi2011, Wu2012, Ghorashi2022, Zhu2022}, and complement recent findings on the band structure modifications by $h$-BN environment~\cite{Long2022} or by encapsulation with other polar materials~\cite{Fumega2023}.

\ColorSection{Electronic properties of 2$h$-BN/BLG and 2$h$-BN/BLG/2$h$-BN heterostructures}
We start with the analysis of the electronic properties of few-layered heterostructures, shown in Fig.~\ref{fig:dft_results}(a), within the density functional theory (DFT) framework.
We neglect the moir\'{e} potential arising from a small lattice mismatch between $h$-BN and graphene layers by setting in-plane lattice constant $a = 2.46$~\AA{} of both components and the interlayer distances to $d = 3.35$ \AA{}.
By fixing the bottom $h$-BN bilayer to be in either AB or AC stacking order, there are 8 non-equivalent configurations of 2$h$-BN/BLG.
For 2$h$-BN/BLG/2$h$-BN with 6 layers in total, the number of possible configurations is larger.
Assuming the same order of B and N atoms in each $h$-BN layer, it is possible to construct 32 configurations.
Only 16 of them are truly independent; for example, the ABCABC configuration has a symmetric partner, ACBACB, related by a combination of the mirror symmetry along the $z$ axis, $\mathcal{M}_z$, and an in-plane shift by a bond length $\tau_{||}$.
Conversely, if the positions of B and N atoms are exchanged in one of the bilayers (which we denote by prime), we can identify 12 stackings connected to their symmmetric partners by the combination of $\mathcal{M}_z \mathcal{M}_x \tau_{||}$ (ABCAC'B' $\longleftrightarrow$ ACBCA'B').
The remaining 8 combinations such as ACACA'C' restore the inversion symmetry, hence their vertical polarization $P_z$ vanishes.
The emerging mirror symmetries in encapsulated heterostructures constrain the value of $P_z$ of symmetry-related stackings to be the same up to the sign~\cite{Wu2021a}.
In 2$h$-BN/BLG heterostructures, there are no spatial symmetries that restrict $P_z$.
Firstly, let us recall the electronic properties of pristine AB-stacked $h$-BN and graphene bilayers.
According to our DFT calculations, free-standing AB-stacked 2$h$-BN has an indirect band gap $\Delta E = 4.78$~eV and a polarization per area $P_z / A = -1.026$~pC/m, with $A = 5.2 \cdot 10^{-4}$~\AA$^2$ being the area of the primitive unit cell~\footnote{We point out that a discrepancy between our values for $P_z / A$ and Refs.~\onlinecite{Yasuda2021, Wu2021a, He2022} is due to a smaller interlayer distance used therein. To corroborate our results, in the Supporting Information document we present the calculations of equilibrium interlayer distances obtained at different levels of DFT}. 
The value of $P_z$ strongly depends on the interlayer distance $d$ and can be further enhanced by decreasing $d$, for example, by applying out-of-plane compressive strain.
On the other hand, BLG is inversion symmetric and gapless with a touching of parabolic bands at the $K$ point.
When inversion symemetry is broken--either by adding extra layers comprising a non-centrosymmetric composite system or by electric field--the BLG develops a gap~\cite{McCann2006, Oostinga2008, Zhang2009}.
The gap opening mechanism in BLG due to sublattice asymmetry has been widely discussed in the literature~\cite{Zhang2011, Tang2017}, but we want to highlight the effect of structural details on the resulting gap.
In Fig.~\ref{fig:dft_results}(b), we show the band structure close to point $K$ of two representative configurations, ABCABC and ACBA.
We found that the band gap $\Delta E$ in graphene/$h$-BN multilayers can vary from $\lesssim$1~meV to 62~meV, whereas $P_z / A$ ranges from $\sim$0.002 to 4.3~pC/m (see Supporting Information for the tabulated results for all studied configurations).
For 2$h$-BN/BLG/2$h$-BN configurations, in which one or more layers have vertically aligned atoms of the same type (below, denoted as primed) $\Delta E$ is on average smaller than in non-primed stacks due to a ``weaker'' violation of inversion symmetry.
Overall, sandwiching BLG with polar $h$-BN bilayers leads to a larger gap opening as compared to one-sided systems.
Interestingly, we do not observe any straightforward relation between $\Delta E$ and $P_z$.
In polar $h$-BN multilayers, one expects that the total polarization of stacked polar bilayers (with the polarization pointing in the same direction) would scale linearly with the number of interfaces~\cite{Stern2021, Yasuda2021}.
A similar behaviour is observed in graphene/$h$-BN heterostructures.
We highlight that if non-polar AA' $h$-BN bilayers are used instead, the entire system can still exhibit a gap and finite polarization.
We therefore conclude the local atomic arrangement at the interfaces is at least an important mechanism responsible for gap opening.
\vfill\null

\ColorSection{From Wannier functions to effective TB models}
As shown in Fig.~\ref{fig:dft_results}(c), the higher conductance bands are composed of the $p_z$ orbitals of B atoms, the lower valence bands have clear contribution of the $p_z$ orbitals of N atoms, while four bands around the Fermi level originate from the $p_z$ orbitals of C atoms. 
This clear separation of the $p_z$ orbitals of carbon atoms allows for the construction of the projector Wannier functions basis sets~\cite{Scaramucci15,Pasquier2019}.
The diagonal matrix elements of the resulting \textit{ab initio} Hamiltonian correspond to the on-site energies of the $p_z$ orbitals.
Build-in electric field arising from $h$-BN bilayers is then reflected as an asymmetry of the on-site terms.
To quantify this sublattice symmetry breaking, we introduce the notion of sublattice imbalance $\Delta V_{\mathrm{sublattice, U/L}} = V_{\mathrm{A, U/L}} - V_{\mathrm{B, U/L}}$, that is, a potential difference between A and B sublattices in the same (upper or lower) layer, and the layer imbalance $\Delta V_{\mathrm{layer}} = (V_{\mathrm{A, L}} + V_{\mathrm{B, L}} - V_{\mathrm{A, U}} - V_{\mathrm{B, U}}) / 2$ characterizing a potential difference between two layers. 
Larger values of $\Delta V_{\mathrm{layer}}$ lead to a stronger modification of the BLG spectrum.
Once the on-site energies from wannierization are obtained, we use them for the construction of an electrostatic potential distribution function $V(\mathbf{r})$, which is then added to a simple nearest-neighbour TB model of AB-stacked BLG
\begin{equation}
    H = \begin{pmatrix}
    V(\mathbf{r}) & t_{||} & 0 & 0\\
     t_{||} & V(\mathbf{r}) & t_{\perp}  & 0 \\
     0 &   t_{\perp} & V(\mathbf{r}) &   t_{||} \\
     0 & 0 & t_{||} & V(\mathbf{r})
    \end{pmatrix}.
\label{eq:hamiltonian}    
\end{equation}
The in-plane hopping $t_{||}= -2.8$~eV and interlayer hopping $t_{\perp} = -0.3$~eV of the Wannier Hamiltonian are consistent with previous reports~\cite{McCann2013}.
To accurately mimic the experimental setup, we impose on $V(\mathbf{r})$ i) a three-fold rotational symmetry $C_{3v}$ along the $z$ axis, and ii) constant value in the triangular domains except the smooth domain boundaries of 4 nm thickness~\cite{Woods2021}.
\begin{figure*}[!htbp]
    \centering
    \includegraphics[width=0.95\textwidth]{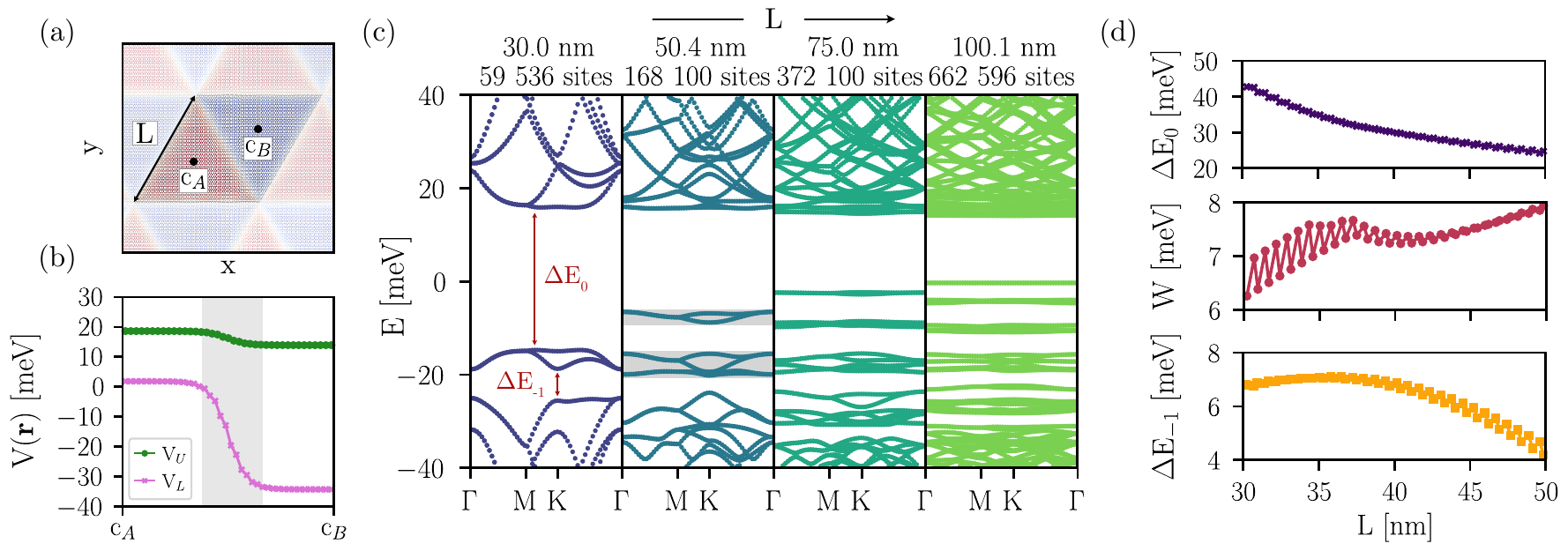}
    \caption{Electronic properties of the BLG superlattices in a polar environment. (a) Electrostatic potential distribution $V(\mathbf{r})$ showing the triangular domains. (b) Corresponding potential cross sections for lower ($V_L$) and upper ($V_U$) layers along the line connecting the centers of triangular domains. Grey shading indicates the domain boundary width. (c) Band structure evolution along $\mathrm{\Gamma}$-$M$-$K$-$\mathrm{\Gamma}$ path at different superlattice periodicities $L$ for the on-site energies corresponding to the ABCABC stacking. At $L = 50$ nm, we observe two subsets of flat bands ($E \sim - 8$ and $E \sim - 18$ meV) decoupled from the remote bands. Upon increasing $L$, a sequence of flat-band manifolds develops at the Fermi level. (d) Variation of the energy gaps above and below the first flat band manifold ($\Delta E_0$ and $\Delta E_{-1}$, respectively), together with its bandwidth $W$, as a function of $L$. From $L =30-50$~nm, the midgap $\Delta E_0$ decreases by $\sim$20~meV, whereas $W$ and $\Delta E_{-1}$ do not change by more than 3~meV.}
    \label{fig:supercell_bands}
\end{figure*}
In Figs.~\ref{fig:supercell_bands}(a,b), we show a distribution of $V(\mathbf{r})$ constructed using the on-site terms for the ABCABC configuration and the potential cross sections along the line connecting the centers of the domains, respectively.

\ColorSection{Flat-band manifolds}
Once parameters dependent on the local structure and composition are clarified, we will address the effect of the superlattice period $L$ as the main tuning knob of the discussed heterostructures. 
From now on, we will focus on the ABCABC stacking configuration characterized by the largest energy gap and $\Delta V_{\mathrm{layer}} = 32$ meV. The results of the wannierization for other heterostructures are presented in the Supporting Information.
In Fig.~\ref{fig:supercell_bands}(c), we show the band structure evolution as a function of the superlattice period $L$ along the high-symmetry path in the mini-Brillouin zone (mBZ).
The spectral symmetry with respect to $E = 0$ is broken due to the asymmetry between sublattices in the two layers.
From $L = 20$~nm (not shown here), a set of two energy bands starts to become disconnected from the rest of the occupied bands.
At approximately $L = 30$~nm, the system develops a gap of 43~meV and these two bands (around $E \sim -18$ meV) further shift in energy.
This subspace is stable in a broad range of values of $L$ as shown in Fig.~\ref{fig:supercell_bands}(d).
As $L$ increases, more bands shift towards the center of the spectrum, forming a series of separated flat-band manifolds composed of even number of bands.
To characterize these bands, we compute the flatness ratio $F$, defined as $F = \Delta / W $, where $\Delta$ is the minimal energy separation from other bands and $W$ is the bandwidth. 
At $L = 50$~nm, there is a subset of two bands around $E = -8$~meV with $W \sim 8$~meV (and $F =0.5 $) and four bands stack around $E = -20$~meV with $W \sim 5 $~meV ($F=0.3 $).
We emphasize that the flat bands should appear for all unprimed configuration with finite $\Delta V_{\mathrm{layer}}$, but at larger $L$.
For primed configurations, though, the situation is more complex--a parabolic dispersion at the $K$ point is still present at $L = 30$~nm, but then it evolves into a Dirac point close to the charge neutrality (see Supporting Information).
A natural question to ask is whether the observed flat bands are topological, similarly to minimally twisted BLG~\cite{Vaezi2013, Zhang2013, Huang2018, Liu2020} or other bipartite lattice systems~\cite{PhysRevLett.125.266403, Regnault2022}.
For now, we focus on two flat-band subspaces appearing at $L = 50$~nm (shaded in Fig.~\ref{fig:supercell_bands}(c)).
The states have weights on both layers, either in the corners or at the centers of domains (consult Fig.~1 in Supporting Information).
The two-fold degeneracy is a consequence of the fact that the supercell Hamiltonian \eqref{eq:hamiltonian} features the $C_{3v}$ symmetry.
As a consequence of band degeneracy at the $\Gamma$ point, the Wilson loop phase $\theta$ is pinned at zero, resulting in a zero winding.
In order to investigate whether the bands can be gapped by a perturbation, we construct an effective Hamiltonian $H_{\mathrm{eff}}$ in the upper flat bands subspace. 
We start with an experimentally accessible way to lift the degeneracy by an external bias $V_0$, which can be realized by a term proportional to the $\sigma_z$ Pauli matrix.
We found that the states remain degenerate for any \textit{globally} applied realistic electric field: a perturbation $V_0$ of the order of $10$~eV results in sub-meV band splitting.
If we allow for \textit{local} perturbations (that is, adding the $V_0$ to the small subset of atomic sites), we can achieve a splitting of few meV for $V_0$ being of the order of few eV.
We highlight that the exact value of $V_0$ and the number of sites to which the perturbation must be applied is rather fine-tuned.
The general conclusion is that $V_0$ does not change the topology of the bands of interest.

\ColorSection{Conclusions}
In summary, we have discussed an experimentally accessible way to alter the electronic properties of bilayer graphene, and vdW materials in general, by electrostatic proximity effect.
We showed that the electronic structure of bilayer graphene can be altered by the electrostatic proximity to polar $h$-BN, and the relevant vertical polarization $P_z$ is controlled by the stacking of individual layers.
We further identified stacking configurations that result in largest gap $\Delta E$ and polarization $P_z$.
We effectively captured the electrostatic potential pattern arising from polar $h$-BN bilayers by performing wannierization onto the $p_z$ orbitals of carbon atoms and used the on-site energies to build TB Hamiltonians for modeling superlattices of relevant periodicities.
By analyzing the TB model for the selected ABCABC stacking of the encapsulated heterostructure, we found a series of flat-band manifolds close to the charge neutrality for a broad range of superlattice periods $L$.
The obtained flat bands manifolds have bandwidths comparable to the bands in twisted bilayer graphene at the magic angle ($5-10$~meV~\cite{Cao2018, Cao2018a}).
Moreover, the emergence of flat bands has been observed in TBG that is partially aligned to the underlying $h$-BN substrate~\cite{Wong2023}.
While the flat bands subspaces are topologically trivial in a single-particle picture, we expect that interactions can open a gap in the many-body spectrum without mixing with remote bands.
Similarly to bilayer graphene or transition metal dichalcogenides, strong electron-electron interactions are expected to lift spin and valley degeneracies, therefore inducing topologically distinct phases~\cite{Zhang2013}.
This is particularly important in the light of experimental realization of the quantum anomalous Hall state with Chern number $C = \pm 1$ at 3/4-filling in the magic-angle twisted bilayer graphene aligned with encapsulating $h$-BN~\cite{Sharpe2019, Serlin2020}.
We leave the analysis of correlations in $h$-BN/BLG polar heterostructures for future work.
A lack of the one-to-one correspondence between the polarization and on-site potentials suggests the importance of electric field screening effects~\cite{Tepliakov21}.
Therefore, it is of great interest to generalize the considerations to more complex graphene/$h$-BN multilayer heterostructures, including the coupling to the in-plane polar textures~\cite{Bennett2023, Bennett2023a}.

\ColorSection{Acknowledgements}
We thank Daniele Passerone for fruitful discussions.
This work was partially supported by the Swiss National Science Foundation (grant No.~172543).
The computations were performed at the Swiss National Supercomputing Centre (CSCS) under Projects No. s1008 and s1146.

\ColorSection{Supporting information}
Details of the computational methodology, discussion on the interlayer distances, methods for computing polarization, details of the Wannier function analysis, discussion of the localization of flat-band states, band structures of primed configurations, tabulated results for all studied configurations.
\bibliography{hBN-graphene.bib}
\end{document}